\RequirePackage{ifpdf}
\ifpdf 
\documentclass[pdftex]{sigma}
\else
\documentclass{sigma}
\fi

\begin{document}

\allowdisplaybreaks

\renewcommand{\thefootnote}{$\star$}

\renewcommand{\PaperNumber}{004}

\FirstPageHeading

\ShortArticleName{Remarks on Multi-Dimensional Conformal Mechanics}

\ArticleName{Remarks on Multi-Dimensional Conformal Mechanics\footnote{This paper is a contribution to the Proceedings of the XVIIth International Colloquium on Integrable Systems and Quantum Symmetries (June 19--22, 2008, Prague, Czech Republic). The full collection
is available at
\href{http://www.emis.de/journals/SIGMA/ISQS2008.html}{http://www.emis.de/journals/SIGMA/ISQS2008.html}}}

\Author{\v{C}estm\'{\i}r BURD\'{I}K~$^\dag$ and Armen NERSESSIAN~$^{\ddag\S}$}

\AuthorNameForHeading{\v{C}. Burdik and A. Nersessian}

\Address{$^\dag$~FNSPE, Czech Technical University in Prague
Trojanova 13, 120 00 Prague 2,
Czech Republic}
\EmailD{\href{mailto:burdik@kmlinux.fjfi.cvut.cz}{burdik@kmlinux.fjfi.cvut.cz}}

\Address{$^\ddag$~Artsakh State University,  5 M.~Gosh Str., Stepanakert,  Armenia}
\Address{$^\S$~Yerevan State University,  1 A.~Manoogian Str., 0025, Yerevan, Armenia}
\EmailD{\href{mailto:arnerses@yerphi.am}{arnerses@yerphi.am}}

\ArticleDates{Received October 30, 2008, in f\/inal form January 10,
2009; Published online January 12, 2009}

\Abstract{Recently,  Galajinsky, Lechtenfeld  and Polovnikov
proposed an elegant group-theoretical transformation of the
generic conformal-invariant mechanics to the free one. Considering
the classical counterpart of this transformation, we relate this
transformation with the Weil model of Lobachewsky space.}


\Classification{70H33; 70H06}

 Since the middle of the seventies, after
\cite{fubini}, much attention paid, in the f\/ield-theoretical
literature, to the simple one-dimensional mechanical system
 given by the Hamiltonian
 \begin{equation}
   H =\frac{{p}^2}{2}+\frac{g^2}{2x^2}.
\label{h}
\end{equation}
The reason was that this Hamiltonian, together with the generators
\begin{equation}
D=px,\qquad K=\frac{x^2}{2} \label{dk}
\end{equation}
 forms the
conformal algebra $so(1,2)$ with respect to the canonical Poisson
brackets $\{p, x\}=1$: \begin{equation} \{  H , D\}=2 H  ,\qquad\{
H , K\}=D,\qquad  \{ K, D\}= -2K. \label{ca} \end{equation} Here
$D$ is the dilatation and $K$ is the conformal boost. It was also
observed, that the action functional of this system is invariant
under conformal transformations, up to total derivative term.
Thanks to this observation the above system  presently known  in
literature as ``one-dimensional conformal mechanics''. Notice,
however, that this system   is conformal invariant in the
f\/ield-theoretical context only, since its action functional
possesses a conformal symmetry provided that the time
reparametrizations are admitted. While the Hamiltonian (\ref{h})
is not invariant under conformal transformation, and the
generators $D$, $K$ do not constitute the conserved quantities,
see~(\ref{ca}). The matter is that in classical  mechanics the
invariance transformations do not admit the time
reparametrizations. This seeming contradiction  can be explained
as follows: to admit the time reparametrizations,  the initial
{\it nondegenerated} Lagrangian should be replaced  by the {\it
degenerated} reparametrization-invariant one. The former
Lagrangian def\/ines  unconstrained Hamiltonian system with
two-dimensional phase space, and the latter one def\/ines the
 constrained Hamiltonian system with four-dimensional phase space,
where the role of additional coordinate plays the  time parameter,
and  the constraint plays  the role of Hamiltonian. The latter
system  is invariant under conformal transformations, given by the
above generators properly extended by the time-dependent terms.
Reducing the system to the nondegenerate one, we arrive to the above
presented one-dimensional  ``conformal mechanics". For the
higher-dimensional mechanical systems
the conformal invariance the same explanation.

One-dimensional conformal mechanics (\ref{h}) can be considered as a two-particle Calogero model
(with the excluded center of mass),  one-dimensional multi-particle integrable
system with in\-ver\-se-square interaction \cite{calogero}
 \begin{equation}
{\cal H}_{\rm Calogero}=\sum_{i=1}^{d+1}\frac{p^2_i}{2}
+\sum_{i<j=1}^{d+1}\frac{g^2}{(x^i-x^j)^2},\qquad \{p_i, q^j\}=\delta^j_i. \label{calog}
\end{equation}
The quantum Calogero model is def\/ined by the above Hamiltonian, where the classical momenta~$p_i$ are replaced by the respective operators
${\widehat p}_i=-\imath\partial/\partial x^i$, and the coupling constant $g^2$ is replaced by $g(g-1)$.
This model has attracted much attention due to numerous
applications in a wide area of physics, as well as due to a rich
internal structure (see, e.g., the recent review~\cite{polychronakos} and references therein).
 Already in the pioneering paper \cite{calogero} it was observed
 that the spectrum of the Calogero model with
an additional oscillator potential is  similar to the spectrum of
free $N$-dimensional oscillator. It was claimed there that a
similarity transformation to the free oscillator system may exist,
at least, in part of  the Hilbert space. However,  this
transformation was written explicitly only decades  after
\cite{hindu}. Later on Galaginsky, Lechtenfeld and Polovnikov found
the unitary transformation, def\/ined  by the
 $SU(1,1)$ group element \cite{galajinsky}, which maps the
 Hamiltonian of the arbitrary conformal mechanics in the Hamiltonian of free
 particle (further we shall refer it as GLP-transformation). This
observation allowed them to  build not only the ${\cal N}=2$
 supersymmetric Calogero model \cite{n2cal}
 but the ${\cal N}=4$ superconformal
Calogero model as well \cite{galaj4}. In~\cite{anton} GLP-transformation
 was extended to  the higher-dimensional many-body systems
 with conformal Galilean symmetry.
So any conformal mechanics can
be mapped on to the free system but the prize is the nonlocality of
 transformation suggested in~\cite{galajinsky}.
  An important and nontrivial  feature of the GLP-transformation is the presence of quantum corrections in the transformed generator
 of conformal boosts: these corrections provide the possibility to built ${\cal N}=4$ superconformal
Calogero.

Recently, one of the authors (A.N.), in collaboration
with T.~Hakobyan,
 proposed a simple classical analog of the decoupling  transformation  for
the one-dimensional conformal mecha\-nics~(\ref{h}) (and for its
${\cal N}=2k$ superconformal extension)~\cite{lobach}.  Namely,
they parameterized the  phase space of the one-dimensional
conformal mechanics by the single complex coordinate
\begin{equation*} { w}=\frac{p}{x}+\frac{\imath g}{x^2}, \quad {\rm
Im}\;{ w}>0: \qquad \{{ w},\bar{ w}\}=-\frac{\imath}{g}\left({
w}-\bar { w} \right)^2. 
\end{equation*} In other words,
the phase space of the system was identif\/ied with the Klein model
of the Lobachevsky plane.

In this   parametrization,  the $so(1,2)$ generators (\ref{h}), (\ref{dk})
def\/ine the Killing potentials (Hamiltonian generators of the
isometries of the K\"ahler structure) of the Lobachevsky plane given
by the following metric and potential
\begin{equation} ds^2=-\frac{g d{ w} d\bar
{ w}}{({\bar w}- { w})^2},\qquad \mathcal{K}=g\log \imath ({\bar w}-
{ w}). \label{klein}
\end{equation}
Explicitly,
\begin{equation}
 H=\imath g\frac{{ w} {\bar w}}{{ w}-{\bar  w}},
\qquad
D=\imath g\frac{{ w}+\bar{ w}}{{ w}-\bar { w}},
\qquad
K=\imath g\frac{1}{{ w}-\bar { w}}.
\label{uk}
\end{equation}
The K\"ahler structure  (\ref{klein}) is invariant under the
 discrete transformation
\begin{equation} { w}\to -\frac{1}{  w}, \label{simt} \end{equation}
whereas the Killing
potentials (\ref{uk}) transform s follows:
 \begin{equation*}
 H\to  K,\qquad K\to H,\qquad D\to -D.
\end{equation*} So in the one-dimensional case everything
is nice and local. However hard to believe, this transformation
can be extended to the $d>1$-dimensional conformal mechanics in a
local (i.e. canonical) way. The discussion of this and related
issues is the purpose of the present note.

For this purpose we, at f\/irst, introduce an appropriate ``radial''
coordinate and  conjugated momentum, so that the higher-dimensional
system looks very similar to the one-dimensional conformal
mechanics. In that picture, the rest (``angular'') of the degrees of
freedom are packed in the  Hamiltonian system on the
$(d-1)$-dimensional sphere: it  replaces the coupling constant $g^2$
in the one-dimensional conformal mechanics, and def\/ines the constant
of motion of the is initial conformal mechanics. Exploiting this
observation we build, by the use of textbook integrable systems on
the spheres, some examples of integrable conformal mechanics. Then
we relate, the radial part of the $d$-dimensional conformal
mechanics with the Klein model of the Lobachevsky space, which is
completely similar to the above-presented one-dimensional case.
However we f\/ind, that an analog of  (\ref{simt}) transformation
necessarily changes the ``angular part'' of the Poisson brackets of
the decoupling, i.e., is noncanonical one.

Let us consider the $d$-dimensional conformal mechanics def\/ined by
the  following Hamiltonian and symplectic structure:
\begin{gather*}
\omega={d{\bf p} }\wedge {d{\bf r}},\qquad {\cal H}=\frac{{\bf
p}^2}{2}+V({\bf r}),\qquad {\rm where}\quad {\bf r}\cdot\nabla V({\bf
r} )=-2V({\bf r}). 
\end{gather*}
 This Hamiltonian together with the generators
\begin{equation*}
D={\bf p}\cdot{\bf r},\qquad  K=\frac{{\bf r}^2}{2} \label{dk1}
\end{equation*}
 forms the
conformal algebra $so(1,2)$ (\ref{ca}). Here $D$ def\/ines the
dilatation and $K$ def\/ines  the conformal boost, ${\bf
r}=(x^1,\ldots, x^d)$, ${\bf p}=(p_1,\ldots, p_d)$, and the Poisson
brackets are assumed to be canonical $\{p_i,x^j\}=\delta^j_i $.

Extracting the radius $r=|{\bf r}|$  we can represent the above
generators  in the following form:
\begin{equation} {\cal D}=p_r r ,\qquad {\cal
K}=\frac{r^2}{2},\qquad {\cal H}=\frac{{p}^2_r}{2}+
\frac{I}{r^2},\qquad I\equiv \frac{{\bf J}^2}{2}+ U,\qquad U\equiv r^2V({\bf r}).
\label{so2}
\end{equation}
 Here  $p_r={\bf p}{\bf r}/r $ is the momentum
conjugated to the radius  $\{p_r, r\}=1$.

It is seen that the
function
 $U$ obeys the condition
\begin{equation*}
{\bf r}\cdot \nabla U =0.
\end{equation*} Introducing some  spherical coordinates $\phi^\alpha $ and
their canonically conjugated momenta $\pi_\alpha $, we get, that
the function $U$ is independent of the radius
\begin{equation*}
\frac{\partial U}{\partial r}=0.
\end{equation*} So, the function
${I}(u)$, $u^a=(\pi_\alpha, \phi^\alpha )$ def\/ines  {\it the
constant of motion} of the system \begin{equation*} \{{\cal H},
I(u)\}=0. \end{equation*} Hence, the conserved quantity $I$ def\/ines
the motion of particle on the $(d-1)$-dimensional sphere, in the
presence of potential $U(\phi )$, so that its phase space is the
cotangent bundle of $(d-1)$-dimensional sphere, $T^*S^{d-1}$.
Parameterizing $T^*S^{d-1}$ by the canonical variables $u^a$, we
can represent this Hamiltonian system
 in the form
 \begin{equation}
 \left(T^*S^{d-1},\; \omega_0=d\pi_\alpha \wedge d\phi^\alpha,\; I(u)=\tfrac{1}{2}{g^{\alpha\beta}(\phi)\pi_\alpha \pi_\beta}
 +U(\phi)\right ).
 \label{hs}
 \end{equation}
 where $g^{\alpha\beta}(\phi)$ is the inverse metrics of the $(d-1)$-dimensional sphere.
It is clear, that  the integrable $d$-dimensional
  conformal mechanics def\/ines the integrable $(d-1)$-dimensional system on the sphere, and vice versa.
It deserves to be mentioned that the generators of conformal algebra
obey the equality:
\begin{equation} 4{\cal H}{\cal K}-{\cal D}^2= 2I.
\label{ml}
\end{equation}

Let us try to solve the equations of motion of the conformal
mechanics in the parametrization given by (\ref{so2}) and
(\ref{hs}). Keeping in mind that $p_r=\dot r$ and f\/ixing the values
of the constants of motion
 ${\cal H}=E$ and $I=I_0 $, we can immediately f\/ind the evolution of
$r$,
\begin{equation*}
t=\int\frac{dr}{\sqrt{2E-\frac{2I_0}{r^2}}}\quad \Rightarrow \quad r^2=\left\{\begin{array}{cc}
\displaystyle 2Et^2+\frac{I_0}{ E}&{\rm for }\;\; E\neq0,\vspace{1mm}\\
\displaystyle 2\sqrt{-2I_0}t &{\rm for }\;\; E=0
\end{array}
\right. 
\end{equation*}
This procedure is completely to the one performed in the Landau and Lifshits textbook on classical mechanics \cite{ll}
for the one-dimensional systems and for the three-dimensional spherically symmetric
systems. The analyzes of the dynamics  of the presented ef\/fective system on from the initial conditions is also similar to those
presented in \cite{ll}.
  Particulary, for  $I_0>0$  we get the one-dimensional system with ef\/fectively repulsive force, and the energy $E$
  takes positive values only. The case  $I_0<0$
 leads to the one-dimensional system with with the ``ef\/fectively attractive force'', and the energy could takes both positive
 and negative values.
In the latter case the range of time could be f\/inite, with corresponds  to the ``falling on the center'' phenomenon.

The evolution of the angular part of the
system is def\/ined by the equations
\begin{equation*}
\frac{du^a}{dt}=\{u^a, {\cal
H}\}=\frac{\{u^a, I(u)\}}{r^2(t)}\quad \Rightarrow\quad
\frac{du^a}{dT}=\{u^a, I\},
\end{equation*}
where $T$ is the new evolution
parameter
\begin{equation*} T= T(t)\equiv\int\frac{dt}{r^2(t)}
\end{equation*}
Hence,
supposing that (\ref{hs}) is the integrable system and we have the
solutions of its equations of motion, $u^a= F^a(T)$,  we can
immediately write down the equations of motion of the initial
conformal mechanics
\begin{equation*}
u^a(t)=F^a(T(t)).
\end{equation*} So with the
appropriate time reparametrization depending on the initial
conditions of the system, we can explicitly integrate its equations
of motion.

Let us represent some examples of the systems of that sort.

\begin{example}
  Adding an arbitrary constant $\kappa $ to the  Hamiltonian of the free particle on $S^{d-1}$
we get the system on $R^{d}$ with the potential \begin{equation}
V=\frac{\kappa}{r^2}. \label{kappa}\end{equation}
 More generally, since the
addition of the constant to the initial Hamiltonian does not change
its integrability property, we can conclude that the adding this
additional term (\ref{kappa}) to the initial conformal mechanics also
preserves the integrability property.

The above-presented  system is the simplest and most known example
of conformal mechanics.
 Nevertheless,   minor modif\/ication of the initial system yields less trivial examples of
 conformal mechanics.

For  example, one can  consider the free particle on $S^2$  moving
in the presence of a constant magnetic f\/ield. This system is still
$so(3)$ invariant. The respective three-dimensional system  describe
the particle moving in the f\/ield of the Dirac monopole.

Similarly, the free-particle moving on $S^4$ in the presence of the
BPST instanton, posses\-ses~$so(5)$ symmetry. The respective
f\/ive-dimensional conformal mechanics describes the particle on~$R^5$
interacting with Yang monopole.
\end{example}

\begin{example}  The $(d-1)$-dimensional spherical (Higgs) oscillator  is def\/ined (up to constant) by the
potential \cite{higgs}
\[
V_{\rm Higgs}=\frac{{\omega^2}\tan^2\theta}{2}.
\]
  It has $(d-1)^2$ constants  of motions
 def\/ining the nonlinear deformation of $su(d-1)$ algebra.
The corresponding conformal mechanics is given by the potential
\begin{equation*} V_{\rm conf}=\frac{\omega^2}{2x^2_{d}}
+\frac{\omega^2}{2r^2}, \end{equation*} and has $(d-1)^2+1 $
constants of motion. Note that this system loses its exact
solvability property in the
 presence of the monopole (for $d=3$) and BPST instanton (for $d=5$) f\/ields.
\end{example}

\begin{example}
The $(d-1)$-dimensional spherical Coulomb system given by the
potential $ V_{\text{Sch-Coul}}{=}$ ${\gamma}\cot\theta $
 was suggested by Schr\"odinger \cite{sch}.
It has $(d-2)(d-3)/2$ constants of motion def\/ining the nonlinear
deformation of the symmetry algebra of the Coulomb system. So the
respective $d$-dimensional conformal mechanics has $(d-2)(d-3)/2
+1 $ constants of motion and is given  by the potential
\begin{equation*} V=\gamma\frac{x_{d}}{r^2\sqrt{r^2-x^2_d}}. \end{equation*}

Let us mention the existence of the integrable spherical analogs of
the two-center Coulomb system \cite{bogush} of the
 anisotropic  oscillator, as well as the integrable spherical analog of the
Coulomb system in the constant electric f\/ield \cite{vahagn}. By the
use of these systems we can construct the respective integrable
conformal mechanical models.
\end{example}

\begin{example} The $(d+1)$-particle Calogero model
 def\/ined by the Hamiltonian (\ref{calog}) can also be represented
in the proposed form.
Since this system is translation-invariant, one can transit to the center-of mass- coordinate frame,
reducing the initial $d$-dimensional system to the $(d-1)$-dimensional one.
 The two-particle Calogero model results, upon excluding
the center of mass, in the one-dimensional conformal mechanics given
by Hamiltonian (\ref{h}). In the three- and more particle case the
resulted system is more complicated. Splitting its radial and
angular parts  we get the integrable system on the
$(d-1)$-dimensional sphere, interacting with the $(d-1)(d-2)/2$
force centers by the Higgs oscillators law. This centers are located
on the sphere in the special order. For example,  for the
three-particle Calogero system we have three force centers on the
circle, with the $2\pi/3$ angles between them. The four-particle
Calogero model yields the system on the two dimensional sphere, with
the force centers located at the vertices of cuboctahedron~\cite{cuboct}.
\end{example}

Let us also mention the recent paper \cite{krivcal}, where the representation (\ref{so2})
has been used for the construction of
${\cal N}=4$ superconformal extension of the three-dimensional conformal mechanics with
the arbitrary positive function $U(u)$, including, as a particular cases,
 the tree-particle Calogero models of various types.

Now let us consider the analog of decoupling transformation of
one-dimensional conformal mechanics, given by~(\ref{simt}) in the
higher-dimensional case. We shall use the representation of
conformal mechanics given by~(\ref{so2}). As we have already
mentioned, the conserved quantity~$I^2$ def\/ines the motion of
particle on the $(d-1)$-dimensional sphere in the presence of the
potential~$U$, so that its phase space is the cotangent bundle of
$(d-1)$-dimensional sphere, $T^*S^{d-1}$. Parameterizing $T^*S^{d-1}$
by the canonical variables $u^a=(\pi_\alpha, \phi^\alpha )$, we can
represent this Hamiltonian system
 in the form given by~(\ref{hs}).

Similarly to the one-dimensional case \cite{lobach}, we introduce, instead of the
radius $r$ and its conjugated momentum $p_r$, the following complex
variable (we restrict ourselves, for simplicity, to a~particular case
$I>1$):
\begin{equation*} { w}=\frac{p_r}{r}+\frac{\imath \sqrt{2I}}{r^2}\equiv
\frac{{\cal D}+\imath \sqrt {2I}}{2{\cal K}}, \qquad
{\rm Im}\;{w}>0.
\end{equation*}
This complex variable obeys the following Poisson bracket relation
\begin{equation}
\{{ w},\bar{ w}\}=-\frac{\imath}{\sqrt{2I(u)}}\left({ w}-\bar { w}
\right)^2.
\label{k1}
\end{equation}
Taking into account equality (\ref{ml}) we can write
\begin{equation}
{\cal H}=\imath \sqrt{2I(u)}\frac{{ w}\bar { w}}{{ w}-\bar { w}},
\qquad {\cal D}=\imath \sqrt{2I(u)}\frac{{ w}+\bar { w}}{{ w}-\bar {
w}}, \qquad {\cal K}=\imath \frac{\sqrt{2I(u)}}{{ w}-\bar { w}},
\label{uk2}
\end{equation}
Now, completely similarly to the one-dimensional
case, described in the introduction, we can def\/ine  the decoupling
transformation
\begin{equation} w\to -\frac{1}{w}\; :\quad {\cal
H}=\frac{p^2_r}{2}+\frac{I}{r^2}\quad\to\quad {\cal
K}=\frac{r^2}{2}, \label{wt}
\end{equation}
 and ${\cal K}\to -{\cal H}$. Keeping
in mind the expressions (\ref{so2}), let us represent the
transformed coordinate~${\widetilde w}$ as follows
\begin{equation*} {\widetilde
w}\equiv -\frac{\widetilde{r}}{\widetilde{p}_r}+\imath \frac{\sqrt{2I(u)}}{\widetilde{p}^2_r}.
\end{equation*}
In that
case, the transformation (\ref{wt}) yields the following
transformation of the radius and conjugated momentum. In the initial
coordinates this  transformation looks like
\begin{equation} {\widetilde
p}_r=\pm \sqrt{{p^2_r}+{2I}/{r^2}}= \sqrt{ 2{\cal H}},\qquad
{\widetilde r}= \pm \frac{p_rr}{\sqrt{{p^2_r}+{2I}/{r^2}}}=\pm
\frac{{\cal D}}{2\sqrt{{\cal H}}}. \label{can2}
\end{equation}
Hence,
everything looks completely  similar to the one-dimensional case.

Let us  notice that the transformed Hamiltonian ${\cal K}$ does not
contain any information on the specif\/ic system. Also, the obtained
transformation (\ref{can2})  can be written without refereeing
 to the internal structure of the Hamiltonian. The only additional constraint is given by
 (\ref{ml}).
The matter is that in our representation the essential  information
on the system is encoded in the Poisson brackets which depend on the
``spherical'' Hamiltonian $I$ and is explicitly def\/ined by the use of
the equations of motion. Indeed, the Poisson brackets are def\/ined
not only by relation~(\ref{k1}), but also by the following ones:
\begin{equation*}
 \{u^a,u^b\}=\omega^{ab}(u),\qquad \{u^a, w\}= (w -\bar w
)\frac{V^a(u)}{2I},\qquad\{u^a, \bar w\}=( w -\bar w
)\frac{V^a(u)}{2I},
\end{equation*}
where $ V^a=\{u^a, I(u)\}$ are the
equations of motion of the spherical system (\ref{hs}), and $\omega^{ab}(u)$ are
its  the Poisson brackets.

 The symplectic structure of the
conformal mechanics can be represented as follows:
 \begin{equation*}
\Omega=-\imath \frac{\sqrt{2I(u)} d{ w}\wedge  d\bar { w}}{({\bar w}-
{ w})^2} + \frac{(dw+d\bar w)\wedge d\sqrt{ 2I(u)}}{\imath(\bar w- w
)}+ \frac{1}{2}\omega_{ab}du^a\wedge {du^b},
\end{equation*}
  while the local
one-form def\/ining this symplectic structure reads
 \begin{equation*} \Omega=
d{\cal A},\qquad {\cal A}=\imath \sqrt{2I(u)}\frac{dw+d\bar
w}{\imath(w-\bar w)} +A_0(u),\qquad dA_0=\frac{1}{2}
\omega_{ab}du^a\wedge du^b.
\end{equation*}
 It is convenient to  represent these
forms as follows:
 \begin{gather*}
 \Omega=-\imath \frac{\sqrt{2I(u) }d{ w}\wedge
d\bar { w}}{({\bar w}- { w})^2} - \imath\frac{w+\bar w}{\bar
w-w}d\log (w+\bar w )\wedge d\sqrt{2I(u)}+
\frac{1}{2}\omega_{ab}du^a\wedge {du^b}, \\ {\cal A}=-\imath \sqrt{2I(u)}\frac{w+\bar w}{\bar
w-w}d\log (w+\bar w) +A_0(u).
 \end{gather*}
  It
is seen that  transformation (\ref{wt}) does not preserves the
symplectic structure, i.e., it is not a canonical transformation
for the  generic $d>1$-dimensional  conformal mechanics. To make the
transformation canonical, we have to admit the time reparametrizations
and to  mix the ``angular'' coordinates
$u^a$ with the ``radial'' one given by $w$.
This means, that we can't make the decoupling transformation canonical in the strong sense
(assuming preservation of the symplectic structure $d{\bf p}\wedge d{\bf r}$).
 But the decoupling transformation could be extended to the ``false" canonical transformation (in the sense  of \cite{ll}),
 which assumes  the preservation of the contact structure ${\bf p}d{\bf r}-{\cal H}dt$
 and, consequently, of the   Hamiltonian
 equations of motion. The discussion of the dif\/ference between these  def\/initions can be found in \cite{arnold}.

\smallskip

\noindent
 {\bf Remark.} Considering the decoupling transformation of conformal mechanics, we restricted ourself
  by the particular case $I>0$. For the $I<0$ case the discussion  and conclusions are very similar former one.
  We should simply replace $\imath\sqrt{2I}$ by $\sqrt{-2I}$. In that case $w$ and $\overline{w}$ become
  {\it real} variables, and the symplectic structure can not be related with the K\"ahler structure.
  Consequently, the generators (\ref{uk2}) can not be interpreted as a Killing potentials.

  \smallskip

Straightforward  transition
 to the quantum mechanics via canonical quantization is ill def\/ined
 procedure for the presented picture. The relevant quantization
 scheme for the one-dimensional conformal mechanics is the  geometric quantization
 (which is well-def\/ined procedure for the quantization on Lobachevsky
space). Even in this simplest case on quantum mechanical level we
shall get some new peculiarities, when consider the decoupling
transformation. Namely, the
  inversion procedure would change the range of validity of the wavefunction. In other worlds,
  it will def\/ine the unitary transformation for  the part of the Hilbert space of the initial system only.
 One can expect, that in the $d>1$ dimensional case we must deal similar to the one-dimensional case,
 with  the radial part,
  encoded in $w$
coordinate. On the other hand in
 in our construction the main role  is given to  the square root from the ``spherical" Hamiltonian, $\sqrt{I}$.
 But taking the ``square root" from the quantum-mechanical
 operator is dangerous. One can, seemingly, to avoid it for the class of quantum spherical Hamiltonians,
 which are factorisable ones. In that case the operator $\sqrt{I}$ will be just the linear operator in the
   canonically quantized system. This will  yields, obviously, nontrivial quantum corrections, observed in
   \cite{galajinsky}.

\subsection*{Acknowledgments} The work is done in accordance with the
research plan of the Czech Ministry of Education MSM6840770039, and
partially supported by grant of the Artsakh Ministry of Education
and Science and by  NFSAT-CRDF UC 06/07 and  INTAS-05-7928 grants.

\pdfbookmark[1]{References}{ref}
\LastPageEnding
\end{document}